# Enabling photoemission electron microscopy in liquids via graphene-capped microchannel arrays

*Hongxuan Guo,[1,2] Evgheni Strelcov,[1,2] Alexander Yulaev,[1,2,3] Jian Wang,[4] Narayana Appathurai,[4] Stephen Urquhart,[5] John Vinson,[6] Subin Sahu,[1,2,7] Michael Zwolak,[1] and Andrei Kolmakov[1]\**

[1]Center for Nanoscale Science and Technology, NIST, Gaithersburg, MD 20899

[2]Maryland NanoCenter, University of Maryland, College Park, MD 20742

[3]Department of Materials Science and Engineering, University of Maryland, College Park, MD 20742, USA

[4]Canadian Light Source, Saskatoon, SK S7N 2V3, Canada

[5]Department of Chemistry, University of Saskatchewan, Saskatoon, SK S7N 5C9, Canada

[6]Material Measurement Laboratory, NIST, Gaithersburg, MD 20899, USA

[7]Department of Physics, Oregon State University, OR 97331, USA

**Photoelectron emission microscopy (PEEM) is a powerful tool to spectroscopically image dynamic surface processes at the nanoscale but is traditionally limited to ultra-high or moderate vacuum conditions. Here, we develop a novel graphene-capped multichannel array sample platform that extends the capabilities of photoelectron spectro-microscopy to routine liquid and atmospheric pressure studies with standard PEEM setups. Using this platform, we show that graphene has only a minor influence on the electronic structure of water in the first few layers and thus will allow for the examination of minimally perturbed aqueous-phase interfacial dynamics. Analogous to microarray screening technology in biomedical research, our platform is highly suitable for applications in tandem with large-scale data mining, pattern recognition, and combinatorial methods for spectro-temporal and spatiotemporal analyses at solid-liquid interfaces. Using Bayesian linear unmixing algorithm, we were able to discriminate between different X-ray induced water radiolysis scenarios and observe a metastable "wetting" intermediate water layer during the late stages of bubble formation.**

*to whom correspondence should be addressed: andrei.kolmakov@nist.gov.

Electron spectroscopy[1, 2] in liquids aims to boost our understanding of the solid-liquid-gas interface relevant to environmental[3], energy[4, 5], catalysis[6] and biomedical research[7]. The pressure gap between the liquid or gaseous sample and the ultra-high vacuum (UHV) partition of the experimental setup (i.e., the electron energy analyzer) is usually bridged via judiciously designed differentially pumped electron optics[8] in combination with an advanced sample delivery systems [9-12]. Experimental challenges, however, delayed the application of the "photon-in electron-out" imaging techniques to solid-liquid interfaces.

Novel 2D materials such as graphene have recently enabled an alternative, truly atmospheric pressure, "photon-in electron-out" X-ray photoelectron spectroscopy (APXPS) [13-17] via separation of the liquid or gaseous sample from UHV with a molecularly impermeable but yet electron transparent membrane. The subnanometer thickness of these membranes is smaller or comparable to an electrons' inelastic mean free path (IMFP). Thus, the photoelectrons are able to traverse the membrane without significant attenuation while preserving their characteristic energies. The drastic reduction of the complexity of the experimental setup allowed the first scanning photoelectron microscopy (SPEM) measurements to be performed in liquid water through graphene-based membranes[14]. However, focused X-ray beam raster scanning during SPEM chemical mapping impedes real-time or prolonged imaging of dynamic processes and decreases the lifetime of the membranes[18]. Therefore, an implementation of the full field of view (FOV) PEEM imaging, is advantageous due to reduced photons density at the sample and acquisition at video frame rate (see, e.g., Bauer[19] and references therein). Though FOV photoelectron imaging of the dynamic processes and objects, such as working catalysts or live cells, in their native high pressure gaseous or liquid environments was a long-standing scientific goal, the differential pumping approach, so successful in APXPS, resulted only in $\approx 10^{-1}$ Pa of near sample pressures so far when applied to the PEEM setup [20]. The near-sample pressure value was limited mainly by the reduced lifetime of the imaging detector and possible discharge development between the sample and PEEM objective lens. An approach, which surmounts these restrictions, was proposed and tested in Ref [21] was based on an environmental cell consisting of two 100 nm to 200 nm thick $Si_3N_4$ membranes with a liquid layer of micrometer thickness in between [22]. The PEEM images of liquid interior of the cell can only be obtained within water soft X-ray transparency window (hv ≈ 285 eV to 532 eV) and in transmission mode. For that $Si_3N_4$ membrane facing the PEEM objective lens was covered with a thin gold photocathode to convert transmitted X-rays to photoelectrons. These very first feasibility tests were, to the best of our knowledge, the only PEEM measurements of hydrated samples so far. On the other hand, prior PEEM research of buried interfaces revealed that the ultra-violet (UV) excited photoelectrons can be recorded through $SiO_2$ films from the depths exceeding many IMFPs[23]. Therefore, standard PEEM imaging in liquids and dense gases can be feasible, in principle, using photoelectrons, provided that UHV and high pressure environments are separated with a thin enough membrane. The latter possibility has been proven recently in an X-ray (X-)PEEM spectromicroscopy study of thermally induced segregation of nano-bubbles at a graphene-Ir interface filled with high pressure (≈ GPa) noble gases. [24]

Here, we develop a novel, versatile microchannel array (MCA) platform that enables a wide range of photoelectron emission spectromicroscopies in liquids through a graphene membrane using UV or soft X-rays. Unlike the case of the aforementioned PEEM "shadow" imaging of immersed objects in the transmission mode, we were able to collect XAS spectro-temporal data of dynamic processes at the graphene-liquid interface *in operando* and submicron spatial resolution using standard laboratory or synchrotron-based PEEM equipment.

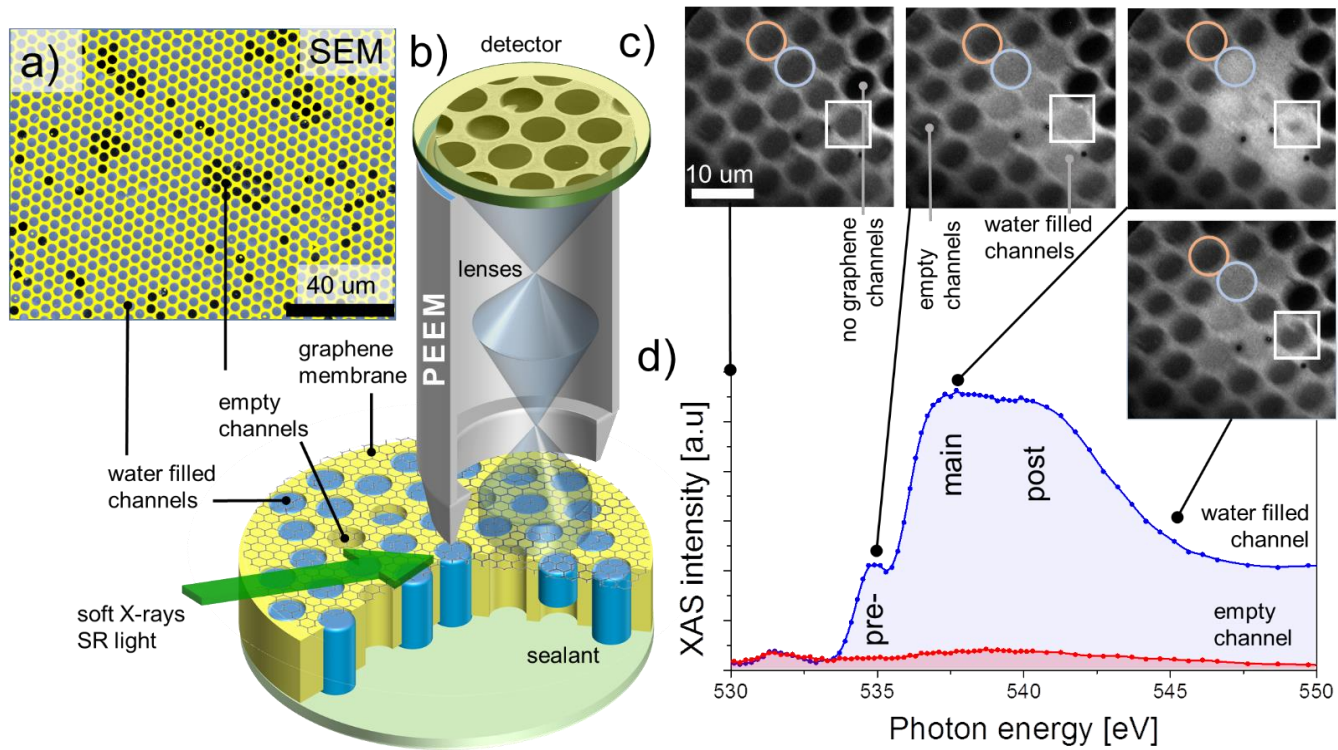

**Figure 1. Multichannel array sample design and experimental setup. a,** SEM (5 keV, color coded) image of water filled graphene capped microchannel sample; the darker channels correspond to the graphene capped but empty channels; **b,** The schematics of the PEEM and liquid cell setups; **c,** PEEM images of the water filled MCA collected at different X-ray energies while crossing the O K-edge; **d,** The resultant XAS spectra collected from different ROIs: water-filled (blue circle and spectrum) and empty (red circle and spectrum) channels. White squares mark the channels that exhibit the dynamic behavior. The spectra were normalized to incident X-ray intensity.

**Multichannel array liquid sample platform for PEEM**

PEEM at liquid-solid interfaces became possible as a result of successful development of a UHV compatible liquid sample MCA platform proposed in Ref [18]. The details of the sample fabrication, liquid filling, and vacuum sealing can be found in the Methods and Supporting Materials sections. Briefly, liquid water was impregnated into the gold coated silica matrix made of an ordered array of ≈ 300 μm deep and ≈ 4 μm wide parallel channels (Fig. 1a). The top of the MCA was covered and isolated from the vacuum with an electron-transparent membrane made of a bilayer graphene (BLG) stack (Fig. 1 b). The bottom of the sample was sealed with a water-immiscible sealant. The water filling factor (the ratio between water filled and empty channels) is routinely in excess of 85 % at the beginning of the experiment in UHV and slowly decays with time. The lifetime of the liquid inside such a sample usually exceeds a few hours and is limited mainly by the graphene quality and interfacial diffusion of water molecules.

The MCA sample containing thousands of water filled micro-channels and capped with BLG was illuminated with monochromatic soft X-rays with an energy between 525 eV and 560 eV, covering the O K-absorption edge ($\approx$ 535 eV). Under this excitation, fast photoelectrons and Auger electrons from the liquid which have the IMFP in excess of the thickness of the capping BLG membrane are able to escape into the vacuum with only minor attenuation[25]. These electrons constitute the total electron yield (TEY) which was used for spatially resolved XAS of the liquid or for spectrally resolved PEEM imaging (Fig. 1b, c). Water containing areas have a sharp characteristic onset in the absorption cross-section around $h\nu \approx 535$ eV and thus can be easily discriminated from the other substrate materials (graphene, Au) which have a flat photoemission background across this energy range.

Figure 1c shows a set of four PEEM images of the BLG-capped water filled MCA recorded at different energies while scanning across the O K-edge. The contrast in these images originates from spatial variations of the local TEY from the Au-coated MCA matrix and the graphene-capped MCA channels. The graphene-capped channels can either be filled with liquid water or be empty. A fraction of the channels does not have a graphene cap and these have the lowest signal in Fig. 1c. As can be seen, the contrast between water filled and empty channels is miniscule below O-K absorption threshold at $h\nu \approx 535$ eV and increases drastically above it. Such sequences of PEEM images constitute a spatial X-ray absorption chemical map and specific regions of interest (ROI) can be designated for site-selective XAS. Figure 1d compares two such XAS spectra collected from two ROIs: water filled (blue) and empty (pink) channels. The empty channels show weak spectral feature at $\approx$ 532 eV characteristic of carbonyl group containing hydrocarbons [26]. These contaminations have been previously observed in XAS of ice and water[27] and in our case can also be due to polymethylmethacrylate (PMMA) residue at the BLG membrane left after graphene transfer[28]. The filled channels, on the other hand, demonstrate an XAS spectrum with pronounced features and a shape typical of liquid bulk water probed via TEY or in transmission detection modes (see reviews [29, 30] and references therein). Such a spectrum is a result or transitions from the strongly localized O 1s core level of water molecules to unoccupied valence orbitals derived from the gas-phase $4a_1$ and $2b_2$ states [31]. The particular feature of these unoccupied orbitals is their *p* character (due to the dipole selection rule) that results in their noticeable directionality and spatial extension far beyond hydrogen atoms. Since hydrogen atoms participate in hydrogen bonding (H-bonding) in water, XAS O K-edge spectra are very responsive to variations in electronic and/or structural environment around the probed water molecule. In good accordance with prior XAS works on liquid water [27, 29], our PEEM-derived XAS spectrum in Fig.1 d has a characteristic pre-edge (≈ 535 eV), main peak (≈ 537.5 eV) features and a post-edge band around ≈ 541 eV. The commonly accepted interpretation of water XAS features assigns these pre-peak and main band to the excitation of water molecules with one broken (or largely distorted H-bond, a so called single-donor (SD) molecules) while the post-edge band corresponds to the molecular environment with strong H-bonds (double donor (DD) molecules) and increased tetrahedrality[31].

It is important to emphasize that the XAS spectrum in Fig. 1 originates from the first few layers of water at the graphene-water interface. This interfacial sensitivity of the through-membrane PEEM spectromicroscopy stems from the attenuating role of the BLG layer, which has low transparency for slow few eV secondary electrons emitted from deeper water layers[32]. Therefore, the bulk-like nature of our spectra indicates that interaction of interfacial water molecules with graphene is very weak and neither the electronic nor the geometrical structure are strongly affected by the graphene. A previous XAS study of interfacial water in contact with gold revealed the significant suppression of the pre-edge peak under similar experimental conditions[3]. Thus, graphene represents a model benchmark material to study interfacial water behavior with PEEM.

**Modelling of the graphene-water interface**

To gain deeper insight in to these differences, we simulate graphene-water structures with all-atom molecular dynamics (MD) simulations and perform calculations of the oxygen K-edge XAS, as described in the Methods and Supplemental Information (SI). As shown in Fig. 2a, b, the structure of the water about 1 nm away from the graphene is already bulk-like.

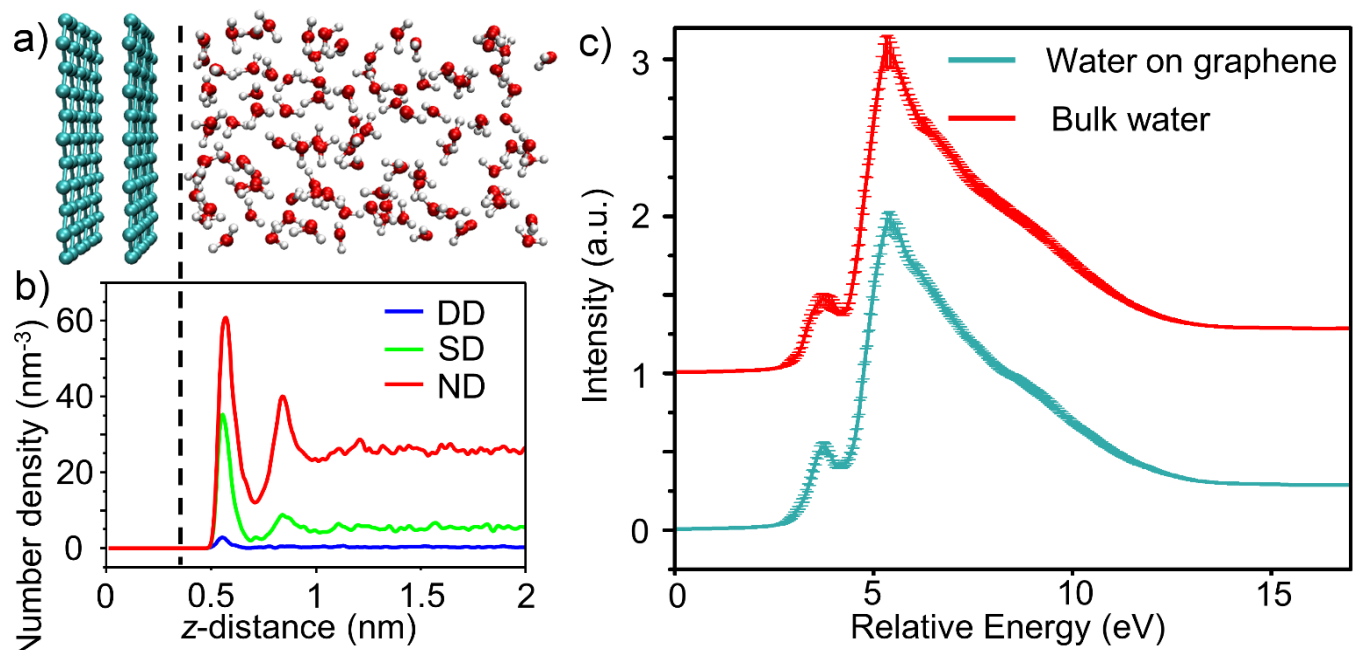

**Figure 2. Water structure, hydrogen bonding, and theoretical XAS spectrum. a,** A snapshot of the bilayer graphene-capped water channel from the MD simulations. The simulation is periodic in all directions with the full unit cell shown in Fig. S1a. **b,** Density of water molecules with differing numbers of hydrogen bonds they donate (DD=double donor, SD=single donor, and ND=non-donor) versus distance from the graphene sheet. **a** and **b** share the same z-axis scale and alignment. The graphene induces a density oscillation in the water, as well as a change in the relative population of different donating species. **c,** The oxygen K-edge XAS spectrum for water at the graphene interface (green) and the spectrum for bulk water (red) as a function of excitation energy. The bulk water spectrum was y-offset for clarity. The presence of graphene does not significantly affect the spectrum of water. This is due to the weak graphene-water interaction: The graphene surface reduces the number of hydrogen bonds, but does not otherwise align water molecules at the surface. Moreover, the core-hole screening by the BLG is not strong enough to suppress the pre-edge peak (as it does with gold). Thus, this peak, as well as the relative location of the main peak are the same for bulk water and water at the graphene interface. The error bars of the XAS curves denote the variance of the mean across different MD snapshots.

The presence of the graphene does, however, induce water density oscillations (see Fig. S2a in the SI) and results in interfacial water losing about 30 % of its hydrogen bonds (Fig. S2b). The latter is reflected in the different proportions of donating species of water near the graphene (Fig. 2b). Despite these changes in interfacial H-bonding and density, the theoretically computed XAS spectrum is similar to one computed for bulk water (Fig. 2c). As with the experimental results, the characteristic pre-edge peak is present in both spectra and of approximately the same relative magnitude compared to the main peak. This is in stark contrast to the aforementioned XAS of liquid water near the gold surface[3], where it was found that the large increase in broken hydrogen bonds (expected to strengthen the pre-edge peak) is overwhelmed by the screening provided by the gold atoms. The highly effective screening of core holes created near the gold surface weakens the core-hole potential, blue-shifting XAS spectrum and reducing the intensity of the lower-energy peaks (see Fig. S3b). On the other hand, the BLG layer does not screen the X-ray induced core hole appreciably, and thus does not suppress the pre-edge peak. In total, for water near a BLG layer, both the structural and electronic effects of the surface are significantly weaker than for water-gold interface, resulting in the similarity of the interfacial and bulk water XAS. We further elaborate on these results in the SI.

**Spectro-temporal evolution of water upon X-ray irradiation**

One of the methodological advantages offered by our MCA sample platform is the automatic collection of a statistical population of geometrically identical objects with variable temporal and spectral behaviors. The latter, in turn, allows for the efficient application of powerful data mining and pattern recognition methodologies. The variations in temporal evolution of water in water-filled channels inside the FOV can already be detected from Figure 1 (e.g., in the channel framed within the square). The whole spectro-temporal 3-dimensional PEEM dataset cannot be directly visualized in 2D plots, and the examples similar to the above (Fig. 1) necessarily require dimensionality reduction via data compression. Therefore, to take advantage of the MCA sample platform and to losslessly compress PEEM datasets, we have employed a multivariate statistical tool – Bayesian linear unmixing (BLU), that has been developed for analysis of hyperspectral imaging datasets.[33, 34] The BLU algorithm reduces a 3D PEEM dataset $Y(x,y,E)$ to a linear combination of position-independent characteristic spectra, $S(E)$, with respective relative abundances, $A(x,y)$: $Y(x,y,E) = S(E) \cdot A(x,y)$. Unlike other statistical tools used for multidimensional data analysis, this method incorporates several built-in constraints that allow for scientifically meaningful interpretation of results. The spectrum at each location therefore can be represented as a linear combination of spectra of individual components in corresponding proportions. The number of spectral components must be provided by the researcher and can be estimated using principal component analysis (PCA) or via under- and oversampling criteria. A detailed description and testing of BLU against PCA, k-means, and other statistical methods can be found elsewhere.[34] The optimal number of components for the dataset shown in Figure 1, was found to be 4 (for details, see SI).

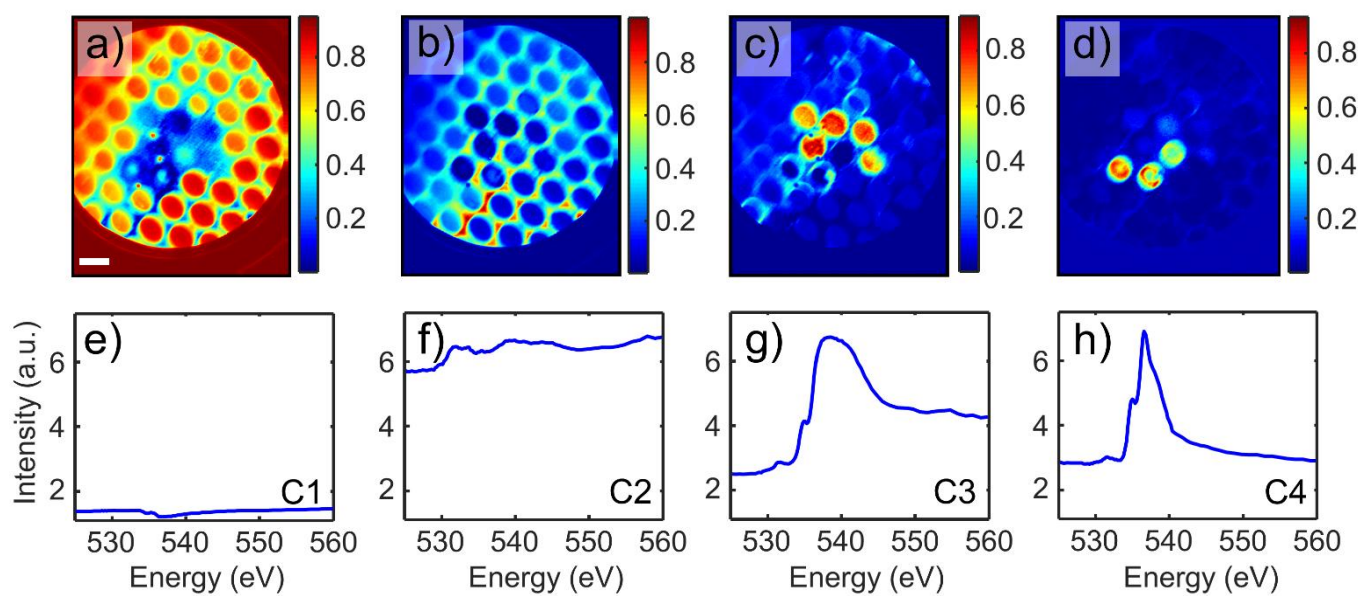

**Figure 3. BLU of a PEEM spectroscopic dataset into 4 components**: Component 1 - panels **a** & **e** empty channels and the frame of the FOV; Component 2 - panels **b** & **f** MCA surface; Component 3 - panels **c** & **g** water-filled static channels; and Component 4 - panels **d** & **h** water-filled dynamic channels. Abundance maps (component intensity as a fraction of unity) and corresponding endmember spectra are shown. The scale bar is 10 μm. Spectra are displayed on the same scale for comparison. Note, that the irradiation intensity was not uniform throughout the sample, the lower left corner being illuminated less than the upper right (cf. panels **a** & **b**).

Figure 3 shows unmixing of a PEEM spectroscopic dataset into 4 components. The signal of component 1 (C1) originates from empty channels of the MCA (also includes the detector frame, see abundance map of Fig. 3a). Its spectrum (Fig. 3e) has low intensity and is almost featureless confirming that there is not any appreciable oxygen absorption in the empty channels. Component 2 highlights the gold-coated surface of MCA (Fig. 3b) that produces a strong background signal with a few weak spectroscopic features (Fig. 3f). This background signal originates from the strong Au X-ray absorption in this energy range, while spectral features are due to carbonyls containing hydrocarbon contaminations of graphene and MCA surface discussed above. We cannot exclude, however, that a part of the spectrum (535 eV to 550 eV) is due to small patches of intercalated water between the graphene membrane and a gold MCA coating. Five water-filled channels in the PEEM FOV produce the strongest signal of component C3 which has well-defined water XAS features and a small contaminant peak at ≈ 533 eV (Fig. 3g). The C3 spatial distribution is seen in the abundance map of Figure 3c. Finally, component C4 is present in three other channels close to the C3 water-filled group (Fig. 3d). Its spectrum is almost identical to that of water in C3 in the range 525 eV to 537 eV; however, beyond this energy, the spectral intensity plummets reaching at 560 eV the same value as at 525 eV. The unusual shape of the C4 spectrum stems from a convolution of a normal water spectrum with its temporal behavior and provides evidence of dynamic changes in the water state taking place underneath the graphene membrane during sample irradiation with X-rays. The dynamic processes activated by intense X-rays or electron beams are well documented[35] and encompass water radiolysis with hydrogen bubble formation[36] followed with water redistribution inside the channel. All of these processes require accumulation of some critical radiation dose to be initialized and lead to a sudden decrease in the intensity of the water TEY signal. The example above illustrates the power of combining the BLU algorithm with X-PEEM technique to recognize the hidden spectro-temporal behaviors in complex systems with mesoscopic spatial resolution.

**Spatiotemporal evolution of water upon X-ray irradiation**

The high spatial and temporal resolution of PEEM allows us to spectroscopically access particularities of soft X-rays induced radiolysis processes in water. In order to explore the dynamics of interfacial water layer up to ultimate bubble formation in detail, we performed time-resolved PEEM measurements by keeping constant the X-rays excitation photon energy (540 eV) and intensity (see movie M1 in SI). The difference between the initial (t=0 s) and final (t=605 s) snapshots of the MCA graphene capped device can be seen in Figure 4a and 4b, respectively. Initially, the device featured some graphene covered but empty channels (black circles in Fig. 4a) and water-filled (light gray circles in Fig 4a) channels. During the first 605 s of sample irradiation, many channels retained water, while others lost water due to radiation induced interfacial bubble formation, evaporation or graphene disruption.

The detailed spatiotemporal evolutions of TEY in corresponding ROIs reveal three different groups of behavior. In the strongly radiolysis-affected water-filled channels, the TEY decreases, forming characteristic step-like drops (e.g. channels 1&2 in Fig. 4 b & d). In the remaining channels, the TEY stays either nearly constant or even increasing by the end of a measurement cycle (channels 4 and 3 of Fig. 4b&d). While a nearly constant TEY indicates that radiolysis products effectively diffuse away from the surface region in these channels, the increase of the TEY in some of the channels is presumably evidence of the buildup of oxygen-rich radiolysis products (e.g. $H_2O_2$) near the graphene-water interface in the strongly confined water volume and/or to graphene oxidation[37].

We now discuss the water-filled channels that are strongly affected by radiolysis. The spatiotemporal map in Figure 4c indicates that TEY evolution is not uniform across the radius in these particular channels. The center of these channels shows the lowest TEY value first at an onset of TEY drop, and then the region of low TEY expanded radially over time until it encompassed the entire channel. This behavior is consistent with radiation-induced bubble nucleation and growth under the graphene membrane. The times at which the electron yield drops and when it reaches the lowest value for a given channel vary widely across the ensemble (see Fig. S5), which reflects the stochasticity in achieving the sufficient supersaturation of $H_2$ to form a bubble inside the irradiated channels. The TEY temporal profiles recorded from the centers of channels 1 and 2 of Figures 4b and 4d have very interesting particularities. The different initial intensity ($I_0$) and the same final intensity ($I_2$) can be explained by different concentration of oxygen-containing species in the probing area and their nearly total absence when the bubble was formed. However, in many cases, the TEY intensity drop is not an instant but has a characteristic intermediate step ($I_1$). Remarkably, the ($I_1$) intensity (normalized to the local irradiation intensity) is nearly the same for the most observed channels. To explain such a "quantized" behavior of TEY intensity evolution, we invoke a simple water multi-layer model (Fig. 4e). The photons at 540 eV energy and grazing angle $16^0$ penetrate ≈ 400 nm deep into the channels in our setup. We presume that Auger electrons with kinetic energies ≈ 500 eV dominate the PEEM TEY intensity due to larger electron attenuation depth within water-BLG stack compared to lower energy secondary electrons[32, 38, 39]. We can, therefore, associate the initial TEY intensity ($I_0$) of the water-filled channel with the PEEM probing depth and the final intensity ($I_2$) with the signal originating from empty channels covered with the graphene membrane. Based on these ultimate values and using standard attenuation formulas[40], an estimate can be made (see SI) of the number of monolayers N contributing to the intermediate step in the TEY intensity drop ($I_1$, Fig. 4d). The numerical value of N depends on the electron inelastic mean free path in water $\lambda_w$ which has not been unequivocally determined yet for our experimental conditions[38, 39, 41]. Assuming $\lambda_w = 2.5$ nm as a conservative water IMFP estimate and 0.25 nm as an effective thickness of a water monolayer[42], Figure 4f presents a map of the estimated number of water layers contributing to the $I_1$-step. As can be seen, the intermediate $I_1$ water state retains between 0.5 to 3 monolayers at the center of individual cells before it disappears completely. Although this number is only a rough estimate it, along with „quantized" behavior of the temporal TEY spectra, suggests the existence of a very thin homogeneous metastable water layer at the surface of the graphene prior to complete evaporation.

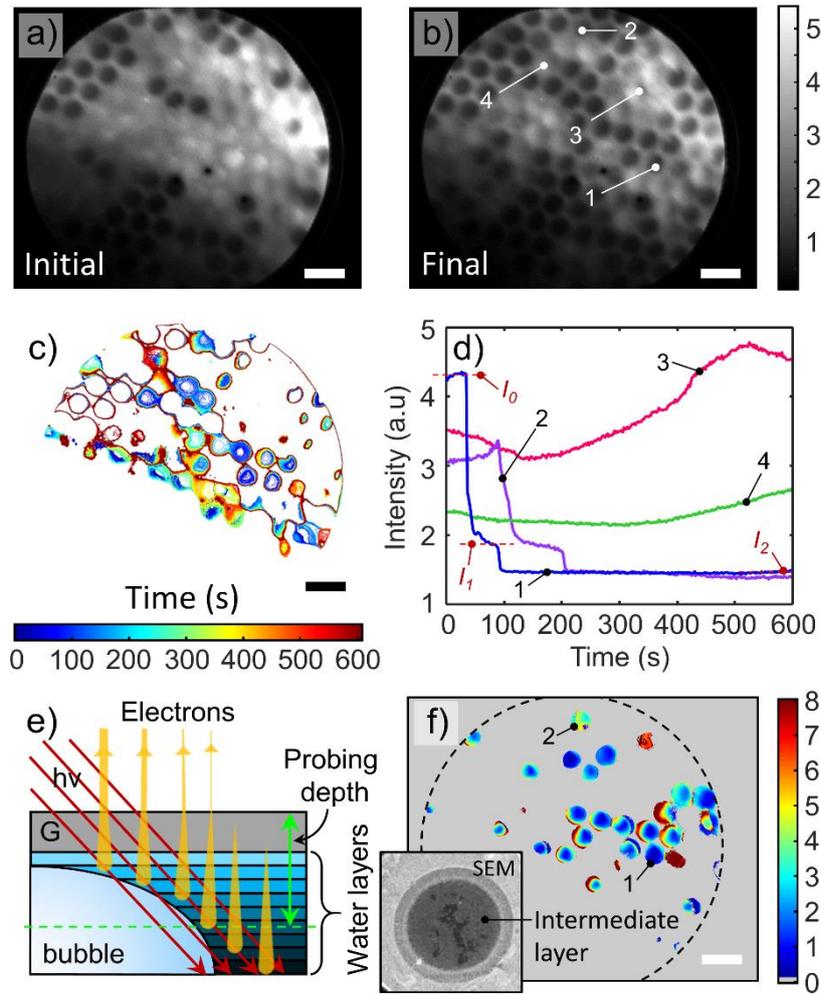

**Figure 4. Spatiotemporal PEEM data analysis**. **a** & **b** PEEM intensity maps (in a.u.) of an MCA graphene capped device as recorded in the initial state and 605 s later at an excitation energy of 540 eV. The temporal behavior of the PEEM signal from the channels can be subdivided into three categories: decreasing XAS intensity, increasing intensity and constant signal intensity. **c,** Contour plot map of the time at which the signal intensity reached a value of 2.2 a.u. showing how the signal drop started at the cell cores and expanded radially outwards. **d**, Intensity vs. time curves averaged over the central region (500 nm × 500 nm) of channels indicated in panel b) and displaying representative behaviors. The decrease in intensity proceeds discretely via formation of steps, until the signal reaches the level of empty channels. The last step is most prominent and is clearly visible for channels 1 and 2. Channels 3 and 4 display fluctuation and increase in intensity that reflect dynamic processes taking place in them. **e,** A schematic of the water-graphene interface model: synchrotron radiation penetrates through the graphene membrane (G) deep into the liquid water, generating photoelectrons, secondary and Auger electrons. Low and intermediate energy electrons become significantly attenuated by cumulative BLG and water layers. Only the fastest electrons with E ≈ 500 eV and IMFP $\lambda_w$ = 2.5 nm (few water layers) contribute to the TEY PEEM signal. **f**, Map of the number of water monolayers that correspond to step $I_1$ (panel d) as estimated from the water-graphene interface model. Dashed circle represents the FOV. Channels 1 and 2 are the same as in panel **b**. Inset: 5 µm × 5 µm screenshot of the SEM movie (2 keV primary energy, in-lens secondary electron detector) of bubble formation inside the water-filled channel at the graphene-water interface. A peripheral multilayer thick liquid water rim surrounds the metastable intermediate $I_1$ layer. Darkest areas correspond to the appeared patches of clean graphene; The scale bars in all images are 10 µm.

The supporting evidence for existence of this intermediate water layer comes from imaging water-loaded MCA devices with scanning electron microscopy (SEM), that is also interface sensitive and provides a higher spatial resolution than PEEM. The SI video M2 and inset in the Fig. 4f demonstrate similar "quantized" behavior of water related electron signal in one of the MCA channels. Despite the difference in radiation doses and image formation mechanisms between the SEM and PEEM, the general radiolysis-induced interfacial-water behavior inside the graphene-covered micro-channels appears to be quite similar. Though the previous reports indicate the formation of the metastable water layers under the confined or low temperature conditions,[43] the understanding of radiation induced stabilization of water layer (for example via dissociative adsorption at the graphene surface[44]) requires further experiments with controlled vapor pressure inside the channels and graphene defectiveness/cleanness. The correlation of this effect with radiation dose makes it plausible that beam induced defects formation and/or chemical modification of the graphene during the radiolysis process could be responsible for this "wetting" phenomenon.

In summary, we developed a novel multichannel, graphene-capped array platform that is UHV compatible and is able to retain liquid samples for hours. The latter, in conjunction with high electron transparency of the bilayer graphene allow us to conduct spectro-microscopy studies of the graphene-water interface with high temporal resolution using standard PEEM instrumentation. The shape of the oxygen K-edge XAS spectra, measured in TEY mode, were similar to bulk water. This result reveals that bilayer graphene does not significantly distort the electronic structure of water in the first few water layers. Our theoretical calculations indicate that this is due to very weak water core-hole screening by the graphene and weak water-graphene interaction. Since the microarray comprises a lattice of identical water filled objects, it is suitable to use this platform in tandem with powerful data mining, pattern recognition, and combinatorial approaches for spectro-temporal and spatiotemporal analysis. Using these algorithms, we were able to discriminate between different scenarios of water radiolysis and detect the appearance of the metastable "wetting" water layer at the later stages of bubble formation. Our work opens up new avenues for investigating electrochemical, catalytic, environmental and other phenomena in liquids using standard (X)PEEM, SPEM, XPS and LEEM setups.

## Methods

*Liquid cell design, graphene transfer, sealing*

MCA is based on a commercial silica based glass matrix used for the fabrication of the multi-channel electron detectors. Before graphene transfer, the top surface of the MCA was metallized with Au (200 nm) / Cr (10 nm) film via sputtering. Monolayer graphene was CVD grown on the surface of a copper foil and coated with a PMMA sacrificial layer. The Cu foil was etched in 200 mol/m$^3$ ammonium persulfate solution. The PMMA/graphene stack was then cleaned in deionized (DI) water and wet transferred onto a monolayer graphene on copper. After drying and annealing of the PMMA/BLG/Cu stack, the metal foil was etched again and after DI water cleaning the PMMA/BLG was transferred on to the Au surface of the MCA. After annealing the PMMA was stripped off by acetone. The acetone was gradually substituted by Isopropyl alcohol and then by DI water at room temperature. In the last step, the water-filled MCA sample was vacuum sealed from the back by UV curable glue or liquid Ga. The latter approach provides a cleaner graphene-water interface.

*PEEM setup*

X-ray PhotoEmission Electron Microscopy (X-PEEM) was conducted at the 10ID-1 SpectroMicroscopy (SM) Beamline of the Canadian Light Source (CLS), a 2.9 GeV synchrotron. The beamline photon energy covers the range from 130 eV to 2700 eV, with a $\approx 10^{12}$ s$^{-1}$ photon flux at the O K-edge (540 eV) and the beamline exit slit size set at 50 μm x 50 μm. The plane grating monochromator (PGM) is able to deliver a spectral resolution of better than 0.1 eV in the measured energy range, and the photon energy scale was calibrated based on samples with known XAS features. The monochromatic X-ray beam was focused by an ellipsoidal mirror down to ≈ 20 μm spot and irradiated on the sample in PEEM at a grazing incidence angle of 16°. The sample is biased at -20 kV with respect to PEEM objective. FOV image stacks (sequences) were acquired over a range of photon energies at the O K-edge. The incident beam intensity was measured by recording the photocurrent from an Au mesh located in the upstream part of the PEEM beamline, and was used to normalize the PEEM data acquired from the sample ROIs. X-PEEM data were analyzed by aXis2000 (http://unicorn.mcmaster.ca/aXis2000.html), and other routine image processing software packages.

*Simulations*

We ran all-atom molecular dynamics (MD) simulations using NAMD[45] with a time step of 1 fs and periodic boundary condition in all directions. The simulation cell consists of 200 water molecules interfacing two parallel sheets of single layer graphene of cross-section 1.2 nm by 1.2 nm with 2 nm of vacuum between them, as shown in Fig. S5a. We use CA type carbon from the CHARMM27 force field and rigid TIP4p water[46]. Van der Waals and electrostatic interactions have a cut-off of 0.6 nm but we perform a full electrostatic calculation every 4 fs via the particle-mesh Ewald (PME) method[47]. To get the production run structures, we minimize the energy of the system in 4 ps and then raise the temperature to 295 K in another 4 ps. Then, we perform a 1 ns NPT (constant number of particles, pressure, and temperature) equilibration using the Nose-Hoover Langevin piston method[48] to raise the pressure to 101,325 Pa (i.e., 1 atm) – followed by 1 ns of NVT (constant number of particles, volume, and temperature) equilibration – to generate the initial atomic configuration. The Langevin damping rate is 0.1 ps$^{-1}$ on all atoms except the carbon atoms (which are fixed during the simulation). The final production run is 0.5 ns of NVT simulation starting with the equilibrated system from which 10 snapshots 50 ps apart were taken for calculating the XAS.

Using the structures from MD, we calculate the oxygen K-edge XAS using the Bethe-Salpeter equation approach implemented within the OCEAN code[49, 50]. Spectra were calculated and averaged over two perpendicular X-ray polarizations in the plane of the graphene. The MD simulation cells were too large to carry out X-ray calculations, so each snapshot was cut down to contain only the single graphene layer surface and the first 128 water molecules placed within a 1.2 nm by 1.2 nm by 4.0 nm box, leaving a vacuum layer of 0.8 nm between the carbon atoms and the periodic image of the water. To account for the short electron inelastic mean free path, we average the contributions of the first 48 oxygen atoms, constituting a depth of approximately 1.0 nm from the shallowest to deepest water molecule below the surface. The bulk spectrum is the result of 5 MD snapshots taken from a 226 water molecule cell.

*Data processing*

The BLU algorithm assumes that a 3D dataset $Y(x,y,E)$ is a linear combination of position-independent endmembers, $S(E)$, with respective relative abundances, $A(x,y)$, corrupted by additive Gaussian noise $N$: $Y(x,y,E) = S(E) \cdot A(x,y) + N$. This method incorporates several built-in constraints that allow physical interpretation of results: the non-negativity ($S_i \geq 0$, $A_i \geq 0$), full additivity and sum-to-one ($\sum A_i = 1$) constraints for both the endmembers and the abundance coefficients. Due to non-negativity of the resulting endmembers S and normalization of abundances, the spectrum at each location can be represented as a linear combination of spectra of individual components in corresponding proportions. The number of spectral components must be provided by the researcher and can be estimated using principal component analysis (PCA) or by under- and oversampling. To solve the blind unmixing problem, the BLU algorithm estimates the initial projection of endmembers in a dimensionality reduced subspace (PCA) via N-FINDR. The latter is a geometrical method that searches for a simplex of maximum volume that can be inscribed within the hyperspectral data set using a simple non-linear inversion. The endmember abundance priors as well as noise variance priors are then chosen by a multivariate Gaussian distribution, where the posterior distribution is calculated based on endmember independence using Markov Chain Monte Carlo. The latter generates asymptotically distributed samples probed by Gibbs sampling strategy. The unmixing error was calculated as $Error(x,y) = \frac{\sum_E (Y(x,y,E) - S(E) \cdot A(x,y))}{\sum_E Y(x,y,E)}$.

**Acknowledgements**

HG, ES, AY and SS acknowledge support under the Cooperative Research Agreement between the University of Maryland and the National Institute of Standards and Technology Center for Nanoscale Science and Technology, Award 70NANB10H193, through the University of Maryland. The high quality graphene was kindly provided by I. Vlassiouk (ORNL, Oak Ridge USA). PEEM measurements were conducted at the Canadian Light Source (CLS) synchrotron radiation facility. The CLS is supported by the Natural Sciences and Engineering Research Council of Canada, the National Research Council Canada, the Canadian Institutes of Health Research, Government of Saskatchewan, Western Economic Diversification Canada, and the University of Saskatchewan. AK, ES thank Prof. A. Hitchcock (McMaster University, Canada) for helpful insights on data processing.

**Author Contributions Statement**

HG, AY and ES contributed to the project equally. HG, AY made and tested the MCA-graphene liquid cells; AK, JW performed the measurements with NA and SU assisting; ES, AK performed analysis of the experimental data; SU, JV, SS, and MZ performed theoretical simulations; AK, ES, MZ, SS and JV co-wrote the manuscript; AK conceived and supervised the project. All authors discussed the results and commented on the manuscript.

**Competing financial interest**

Authors declare no competing financial interests.

ASSOCIATED CONTENT

Supporting Information

The Supporting Information is available free of charge on the

ACS Publications website at http://pubs.acs.org/journal/nalefd

TOC Figure

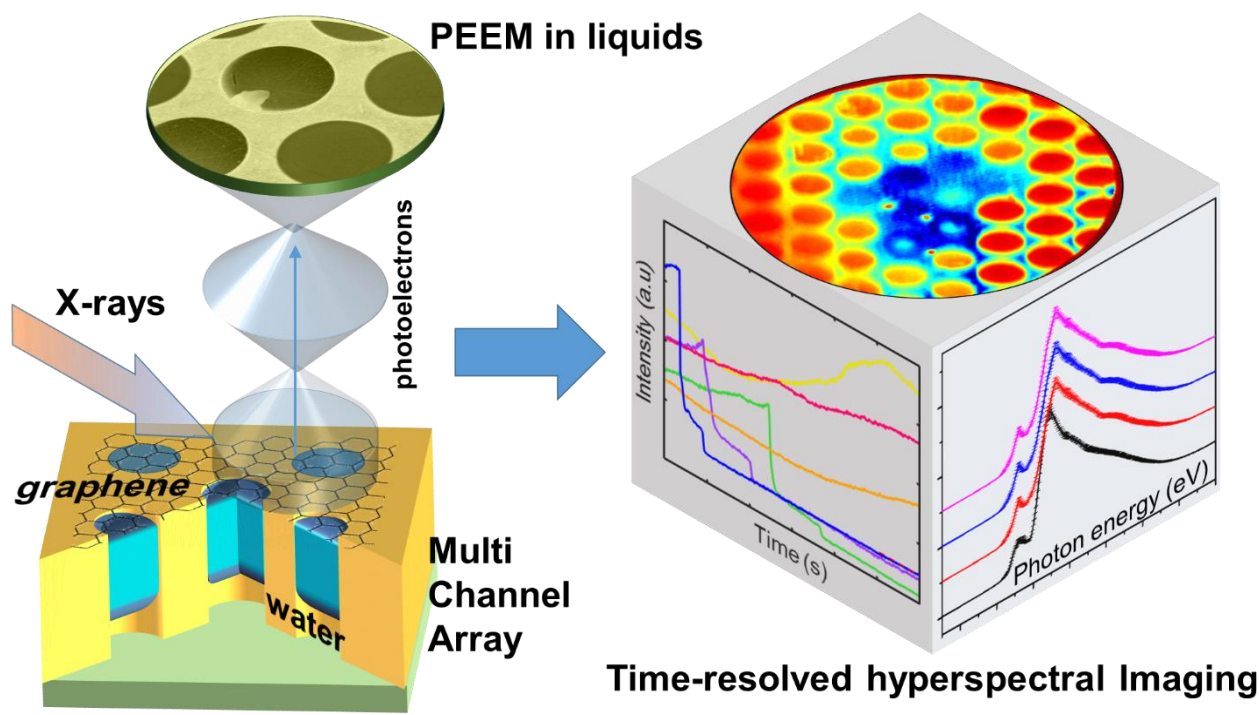